\documentclass[12pt]{article}
\newcommand{\be}{\begin{equation}}
\newcommand{\ee}{\end{equation}}
\newcommand{\beq}{\begin{equation}}
\newcommand{\eeq}{\end{equation}}
\newcommand{\ba}{\begin{eqnarray}}
\newcommand{\ea}{\end{eqnarray}}

\begin{document}
\baselineskip=15.5pt
\pagestyle{plain}
\setcounter{page}{1}


\def\del{{\partial}}
\def\vev#1{\left\langle #1 \right\rangle}
\def\cn{{\cal N}}
\def\co{{\cal O}}
\newfont{\Bbb}{msbm10 scaled 1200}     
\newcommand{\mathbb}[1]{\mbox{\Bbb #1}}
\def\IC{{\mathbb C}}
\def\IR{{\mathbb R}}
\def\IZ{{\mathbb Z}}
\def\RP{{\bf RP}}
\def\CP{{\bf CP}}
\def\Poincare{{Poincar\'e }}
\def\tr{{\rm tr}}
\def\tp{{\tilde \Phi}}

\def\TL{\hfil$\displaystyle{##}$}
\def\TR{$\displaystyle{{}##}$\hfil}
\def\TC{\hfil$\displaystyle{##}$\hfil}
\def\TT{\hbox{##}}
\def\HLINE{\noalign{\vskip1\jot}\hline\noalign{\vskip1\jot}}
\def\seqalign#1#2{\vcenter{\openup1\jot
  \halign{\strut #1\cr #2 \cr}}}
\def\lbldef#1#2{\expandafter\gdef\csname #1\endcsname {#2}}
\def\eqn#1#2{\lbldef{#1}{(\ref{#1})}%
\begin{equation} #2 \label{#1} \end{equation}}
\def\eqalign#1{\vcenter{\openup1\jot
    \halign{\strut\span\TL & \span\TR\cr #1 \cr
   }}}
\def\eno#1{(\ref{#1})}
\def\href#1#2{#2}
\def\half{{1 \over 2}}

\def\ads{{\it AdS}}
\def\adsp{{\it AdS}$_{p+2}$}
\def\cft{{\it CFT}}

\newcommand{\ber}{\begin{eqnarray}}
\newcommand{\eer}{\end{eqnarray}}

\newcommand{\beqar}{\begin{eqnarray}}
\newcommand{\cN}{{\cal N}}
\newcommand{\cO}{{\cal O}}
\newcommand{\cA}{{\cal A}}
\newcommand{\cT}{{\cal T}}
\newcommand{\cF}{{\cal F}}
\newcommand{\cC}{{\cal C}}
\newcommand{\cR}{{\cal R}}
\newcommand{\cW}{{\cal W}}
\newcommand{\eeqar}{\end{eqnarray}}
\newcommand{\tht}{\thteta}
\newcommand{\lm}{\lambda}\newcommand{\Lm}{\Lambda}
\newcommand{\eps}{\epsilon}


\newcommand{\nonu}{\nonumber}
\newcommand{\oh}{\displaystyle{\frac{1}{2}}}
\newcommand{\dsl}
  {\kern.06em\hbox{\raise.15ex\hbox{$/$}\kern-.56em\hbox{$\partial$}}}
\newcommand{\id}{i\!\!\not\!\partial}
\newcommand{\as}{\not\!\! A}
\newcommand{\ps}{\not\! p}
\newcommand{\ks}{\not\! k}
\newcommand{\D}{{\cal{D}}}
\newcommand{\dv}{d^2x}
\newcommand{\Z}{{\cal Z}}
\newcommand{\N}{{\cal N}}
\newcommand{\Dsl}{\not\!\! D}
\newcommand{\Bsl}{\not\!\! B}
\newcommand{\Psl}{\not\!\! P}
\newcommand{\eeqarr}{\end{eqnarray}}
\newcommand{\ZZ}{{\rm \kern 0.275em Z \kern -0.92em Z}\;}

\begin{titlepage}

\begin{center} \Large \bf Quantum Relaxation of the Cosmological Constant ${}^{**}$

\end{center}

\vskip 0.3truein
\begin{center}
R. Jackiw${}^{\,*}\footnote{jackiw@lns.mit.edu}$, Carlos N\'u\~nez${}^{\,
*}$
\footnote{nunez@lns.mit.edu}
and S.-Y. Pi
${ }^{\,\dagger}$
\footnote{soyoung@bu.edu}

\vspace{0.3in}
${}^{*}$ Center for Theoretical Physics, Massachusetts Institute of
Technology \\
Cambridge, MA 02139, USA

\vspace{0.3in}

${}^{\,\dagger}$Department of Physics \\
Boston University, Boston, MA 02215,
USA
\vspace{0.3in}

\end{center}
\vskip 1truein

\begin{center}
{\bf ABSTRACT:}
We describe a mechanism that drives the Cosmological 
Constant to zero 
value. This mechanism is based on the quantum triviality of $\lambda 
\phi^4$ field theory and works in $AdS$ space. Some subtleties of the 
model are 
discussed. 
\end{center}

\vskip2.6truecm
\vspace{0.3in}
${}^{**}$ Einstein Memorial Issue, Physics Letters {\bf A}. 

Presented at the Kummerfest, Vienna, January 2005.
\smallskip

\vspace{0.3in}
\leftline{BUHEP-05-04 }
\leftline{MIT-CPT 3609}
\leftline{hep-th/0502215 }
\smallskip
\end{titlepage}
\setcounter{footnote}{0}


\section{Introduction}
Astrophysical observations, which have been interpreted as
evidence for a cosmological constant $\Lambda$ \cite{Spergel:2003cb}, have moved the
``cosmological constant problem'' from explaining a vanishing value,
$\Lambda=0$, to explaining a non-vanishing but tiny positive
value, (for reviews see \cite{Weinberg:1988cp},\cite{Padmanabhan:2002ji}, 
\cite{Nobbenhuis:2004wn}). In this note we
remain with the original
problem.
We discuss a possible mechanism that could drive $\Lambda$ to zero,
in the  belief that
once a vanishing cosmological constant is secured, raising it to its tiny
but non-vanishing value is a  milder problem.

The mechanism for driving $\Lambda$ to zero, to which here we call
attention,
is a quantum effect
encountered in
the  $\lambda \phi^4$ theory:
while classically $\lambda$ can take any  value, in the quantized theory only
$\lambda=0$ is possible \cite{Wilson:1971}. Of course, a regularized 
quantum field 
theory may posses any value for $\lambda$ but
this value
vanishes as the
regulator is removed.

Before stating the proposal
let us first set the conventions that we shall be using. The metric 
signature will be
mostly minus
$\eta_{\mu\nu}=(+,-,-,-)$.
The Christoffel symbol is given by
\eqn{cris}{\Gamma^\lambda_{\mu\nu}= \frac{g^{\lambda\alpha}}{2}
(\partial_\mu g_{\alpha\nu} + \partial_\nu g_{\alpha\mu} -\partial_\alpha
g_{\mu\nu}     ).}
The Riemann and Ricci tensors are defined as
\eqn{riemann}{R^\mu_{~\nu\alpha\beta} =
 \partial_{\alpha}\Gamma^{\mu}_{\beta \nu} -
\partial_{\beta}\Gamma^{\mu}_{\alpha\nu} +
\Gamma^{\mu}_{\alpha\sigma} \Gamma^{\sigma}_{\beta\nu}  -
\Gamma^{\mu}_{\beta\sigma}  \Gamma^{\sigma}_{\alpha\nu}, \;\;\;\;R_{\mu\nu}=
R^\alpha_{~\mu \alpha \nu}, \,\, R= g^{\mu\nu}R_{\mu\nu}.     }
With these conventions, the equations of motion derived from the 
Einstein-Hilbert action
\eqn{0}{S=-\frac{1}{16 \pi G}\int d^4 x \sqrt{-g} (R - 2 \Lambda),} read
\eqn{eq}{R_{\mu\nu}=   g_{\mu\nu}\Lambda,}
and consequently
\eqn{ricci scalar}{R= 4\Lambda.}
For $\Lambda>0$~($\Lambda<0$) this corresponds to Anti de Sitter, $AdS$ 
(de
Sitter, $dS$) space.
The units of the quantities above are
$[G]= m^{-2}, \;\;[R]=m^2, \;\;[\Lambda]=m^2,\;\; [x]=m^{-1} $.
%
%
\section{The proposal}
We consider the Einstein-Hilbert action
\eqn{ei}{S=-\frac{1}{16 \pi G}\int d^4 x \sqrt{-g} (R- 2 \Lambda),}
and propose that $\Lambda$, which is arbitrary in the classical theory,
will be driven to zero by 
quantum effects.

Owing to the non-renormalizability of (\ref{ei}), it is not possible
for us to asses our proposal convincingly in General Relativity.
Nevertheless, the following calculation supports the proposal. In
eq. (\ref{ei}), rescale the metric tensor as \eqn{metr} {
g_{\mu\nu}= \phi^2 \hat{g}_{\mu\nu},\;\;\;\; ds^2= \phi^2
\hat{ds}^2.} In four dimensional spacetime, this will scale the
volume factor as $\sqrt{-g}=\phi^4 \sqrt{-\hat{g}}$ and the Ricci
scalar  as \eqn{curv4}{R(g)= \phi^{-2} R(\hat{g}) - 6\phi^{-3}
\hat{D}^2\phi.} All 
quantities on the right, including the covariant derivative $\hat{D}$ involve 
the rescaled 
metric $\hat{g}_{\mu\nu.} $ 
\footnote{\renewcommand{\theequation}{\roman{equation}}
\setcounter{equation}{0} If in
d-spacetime dimensions, we choose a Weyl transformation  
as $g_{\mu\nu} = \phi^{2k}\hat{g}_{\mu\nu}$, the Ricci scalar
changes as (see for example \cite{Wald:1984rg}) \eqn{ricciD}{R(g)=
\phi^{-2k} \left( R(\hat{g}) - 2 k (d-1)\phi^{-1} \hat{D}^2\phi  + 
[2 k(d-1)- k^2 (d-1)(d-2)]\phi^{-2} \hat{g}^{\mu\nu}\partial_\mu \phi
\partial_\nu\phi)\right),} and  the Einstein-Hilbert action reads, after 
integrations 
by parts, \eqn{changeD}{\int d^d x \sqrt{|g|} R = \int d^d x
\sqrt{|\hat{g}|}\phi^{k(d-2)} [R(\hat{g}) + 
k^2(d-1)(d-2)\phi^{-2}\hat{g}^{\mu\nu}\partial_\mu\phi
\partial_\nu\phi ].} In the exponent factor, $k$ is set to  $k=\frac{2}{(d-2)}$
by requiring that the scalar kinetic term be conventional. Then eq.
(\ref{ricciD}) implies \eqn{Roman}{\frac{(d-2)}{8(d-1)}\int d^dx
\sqrt{|g|}(R(g) - \frac{8(d-1)}{(d-2)} \lambda) = \int d^dx
\sqrt{\hat{|g|}} [\frac{1}{2} \hat{g}^{\mu\nu} \partial_\mu
\phi\partial_\nu \phi +\frac{(d-2)}{8(d-1)} \phi^2 R(\hat{g})
-\lambda \phi^{2d/(d-2)}].}

This shows that (in any $d>2$) the Weyl invariant action for a conformally
coupled scalar field with self interaction, as on the right 
in (\ref{Roman}), is equal to the Einstein-Hilbert action with a cosmological constant and a 
rescaled
metric, as on the left in (\ref{Roman}). 
This identity is a consequence of Weyl invariance which may be 
used to set $\phi=1 $ on  the 
right hand side of
eq.(\ref{Roman}), therefore achieving the left side.}
\setcounter{equation}{8}
After partial integration, with surface terms ignored, the action
(\ref{ei}) becomes \eqn{action2}{S=-\frac{3}{4 \pi G}\int d^4 x
\sqrt{-\hat{g}}~[\frac{1}{12} R(\hat{g})\phi^2 +
\frac{1}{2}\hat{g}^{\mu\nu}\partial_\mu\phi\partial_\nu\phi -\frac{
\Lambda}{6} \phi^4]. } In order to have canonical units for fields
and couplings, we define 
\eqn{defi}{\varphi=
\frac{\phi}{\sqrt{G}}, \;\; \Lambda =\frac{\lambda}{4 G},} and
finally the action (\ref{ei}) reads, 
\eqn{sfinal}{S= -\frac{3}{4\pi}\int 
d^4 x
\sqrt{-\hat{g}}~ ( \frac{1}{12}R(\hat{g})\varphi^2 +
\frac{1}{2}\hat{g}^{\mu\nu}\partial_\mu \varphi \partial_\nu\varphi
-\frac{\lambda}{4!}\varphi^4 )~.}

Let us now, use (\ref{sfinal}) in two ways. First working in a theoretical laboratory with mini-superspace  variables, we set
$\hat{g}_{\mu\nu}$ to be the Minkowski metric, and retain only the conformal factor $\varphi$. 
Eq.(\ref{sfinal}) shows that 
the field $\varphi$ follows the dynamics of a
$\lambda \varphi^4$ theory, which according to K. Wilson \cite{Wilson:1971}, makes $\lambda$ vanish in the
quantum field theory. Second, we make the physical argument
that it does make sense, for our present day universe, to take a flat background
metric. We are then again left with a $\lambda \varphi^4$
theory, which is suppressed quantum mechanically. 
Guided by these observations, we suppose that Wilson's argument for $\lambda \varphi^4$ theory 
can be extended from 
flat space-time \cite{Wilson:1971} so that it holds for (\ref{action2}) and (\ref{sfinal}).
For example a constant curvature background gives to the $\varphi$ field a mass and changes 
the 
kinetic term, but this should not modify the short distance behavior needed for  
Wilson's argument.

Let us note the issues that arise because of the signs in
(\ref{sfinal}). In the flat limit, we need $\lambda$ in (\ref{sfinal}), to be positive
relative to a positive kinetic term. Thus this relaxation mechanism
works in AdS space. The overall sign of the action in (\ref{sfinal})
is negative, compared to the usual matter action (in our
conventions). This would render the $\varphi$ dynamics unstable if
the field $\varphi$ couples to other matter fields. We evade this
problem by assuming that the dynamics of matter fields is Weyl
invariant, hence independent of $\varphi$. In the Standard Model,
this is true, except for the potential energy in  the Higgs sector; 
but little is certain about Higgs 
dynamics. Indeed, the action of the gauge
fields scales as (the gauge potential does not scale)
\eqn{gaugef}{S_{gauge}=\int d^4 x\sqrt{-g} g^{\mu\nu} g^{\alpha\beta}
F_{\mu\alpha} F_{\nu\beta} \to \int d^4x
\sqrt{-\hat{g}}\hat{g}^{\mu\nu} \hat{g}^{\alpha\beta} F_{\mu\alpha}
F_{\nu\beta}. \;\;} For the fermions, which  scale as
$\psi=\phi^{-3/2}\hat{\psi}$, the action (for a charged fermion) in
curved space is also Weyl invariant. 
\ba
& & S_{fermion}=\int d^4x\sqrt{-g} 
\bar{\psi}\gamma^\mu D_{\mu}(\omega,A)\psi = \int
d^4 x\sqrt{-g} \bar{\psi} \gamma^\mu (\partial_\mu +\frac{1}{4}
\omega_{\mu}^{ab}\gamma_{ab} + A_\mu)\psi \to \nonumber\\
& & \int d^4 x
\sqrt{-\hat{g}}\bar{\hat{\psi}} \hat{\gamma}^\mu \hat{{\cal D}}_{\mu}
(\hat{\omega},A)\hat{\psi}. 
\label{fermions}
\ea

We emphasize that the quantum
suppression of the quartic self coupling (and therefore of the
cosmological constant) is not driven by the perturbative
renormalization group, because perturbation theory is inapplicable
to $\lambda \varphi^4$ theory when the regularization is removed.
Indeed, the perturbative renormalization group 
with the lowest order beta function $c \lambda^2, c=3/16\pi^2$ gives a running
\eqn{running}{\lambda(\mu)= \frac{\lambda(\mu_0)}
{1- c \lambda(\mu_0) \log(\frac{\mu}{\mu_0}) }.} This perturbative
results, holds only for very small (compared to 1) values of
$\lambda$. For the `original' cosmological constant $\Lambda$ this implies 
\eqn{running2}{\Lambda(\mu)= 
\frac{\Lambda(\mu_0)}{1- 4 c \Lambda(\mu_0) G \log(\frac{\mu}{\mu_0}) },}
which does not provide sufficient running from a sizeable
cosmological constant at a high energy $\mu$ to its small value at
$\mu_0= 2.3 K$ in energy units. Perturbation theory, even
when improved by the renormalization group, does not address our
proposal. 

However, one can encounter 
within perturbation theory indications that the 
renormalized $\lambda$ 
must vanish. This plausibility argument for triviality 
of  $\lambda \varphi^4$ is  found within 
the lowest 
order, dimensionally regulated renormalization group analysis 
\cite{Brown:1992cu}. 
Setting 
dimensionality ~~ $\epsilon =( 4-d)$, we recognize that (\ref{running}) 
and (\ref{running2}) 
arise in the limit  $\epsilon \to 0$ from the equality
\beq
\lambda_0 \ = \ 
\mu_1^{\epsilon} 
\frac{\lambda (\mu_1)}{( 1 - \lambda ({\mu_1})\frac{c}{\epsilon})} \ 
= \ \mu_2^{\epsilon} \frac{\lambda (\mu_2)}{( 1 - \lambda 
({\mu_2})\frac{c}{\epsilon})} 
\eeq
where $\lambda_0$ is the bare coupling. 
The theory should be well defined for $d=4 - \epsilon$ with $\epsilon > 
0$. 
To reach physical 4 dimensions we let $\epsilon \to 0^+$. 
But before reaching $ d=4$ we encounter the singularity 
at $\epsilon = c \lambda (\mu)>0$ where, as stressed above, the theory 
should be well defined. This causes 
the bare coupling 
$\lambda_0$ to diverge, So, the problem is avoided by supposing that the 
renormalized 
coupling vanishes, and only a non-interacting theory remains.

The idea then is that a field theory, 
which is not well defined for large values of the 
energy cut-off solves its problems by becoming non-interacting 
when the problematic energy scale is reached. 
Our proposal for the cosmological constant relies on 
such a non-perturbative suppression of the coupling constant.

There
exist numerous studies of triviality of  $\lambda\varphi^4$ theory.
Most of them are motivated by the fact that the Higgs potential in
the standard model should show a behavior like the one described
above. 
 These studies include lattice simulations that support
the result (see for example \cite{Kuti:1987nr} and  references
cited in \cite{Callaway:1988ya} ). Also there are proofs of
triviality for space times with dimension greater than four. In
four dimensions there are arguments that are very suggestive  of
triviality, but no definitive proof \cite{Frohlich:1982tw}.

After formulating this argument, we were reluctant to put 
it forward 
publicly, because of the many lacunae. But then we were delighted and 
encouraged to discover that a similar proposal for suppressing the 
cosmological constant was made by A. Polyakov in
his Oscar Klein Memorial Lecture \cite{Polyakov:2000fk}.

\section{Acknowledgments:} We thank colleagues for discussions and 
useful comments.
The work of R.Jackiw, C. Nu\~nez and S.-Y.Pi was supported by
by funds provided by the U.S Department of Energy (DOE)  under 
cooperative research agreements DE-FC02-94ER-40818 and 
DE-FG02-91-ER-40676.
C. Nu\~nez is also supported by  a Pappalardo Fellowship.

\end{document}